\documentclass[floatfix,a4paper,aps,prl,twocolumn,preprintnumbers]{revtex4} 

\usepackage{epsfig,ulem}
\usepackage{amsmath}
\usepackage{amssymb}
\usepackage{amsfonts}
\usepackage{graphicx}
\usepackage{graphics,subfigure}
\usepackage{float}
\usepackage{booktabs}
\usepackage{color}
\usepackage{geometry}
\usepackage{rotating}

\newcommand{\newc}{\newcommand}

\newc{\lam}{\lambda}
\newc{\eps}{\epsilon}
\newc{\kap}{\kappa}

\newc{\ra}{\rightarrow}
\newc{\ovl}{\overline}
\newc{\lsim}{\stackrel{<}{\sim}}
\newc{\Tr}{{~\rm Tr}}

\begin{document}
  
\vspace*{-0mm}  
  
\title{What if the LHC does not find supersymmetry in the
  $\sqrt{s}=7$~TeV run?}

\author{Philip Bechtle}
\affiliation{Deutsches Elektronen-Synchrotron DESY, Notkestra{\ss}e 85, D-22607 Hamburg, Germany}

\author{Klaus Desch}
\affiliation{Physikalisches Institut der Universit\"at Bonn, Nussallee 12, D-53115 Bonn, Germany}

\author{Herbi K.\ Dreiner}
\affiliation{Physikalisches Institut der Universit\"at Bonn, Nussallee 12, D-53115 Bonn, Germany}

\author{Michael Kr\"amer}
\affiliation{Institute for Theoretical Particle Physics and Cosmology, RWTH Aachen University, D-52056 Aachen, Germany}

\author{Ben O'Leary}
\affiliation{Institute for Theoretical Physics and Astrophysics, W\"urzburg University, Am Hubland, D-97074 W\"urzburg, Germany} 

\author{Carsten Robens}
\affiliation{Institute for Theoretical Particle Physics and Cosmology, RWTH Aachen University, D-52056 Aachen, Germany}

\author{Bj\"orn Sarrazin}
\affiliation{Deutsches Elektronen-Synchrotron DESY, Notkestra{\ss}e 85, D-22607 Hamburg, Germany}

\author{Peter Wienemann}
\affiliation{Physikalisches Institut der Universit\"at Bonn, Nussallee 12, D-53115 Bonn, Germany}

\begin{abstract}
\noindent
We investigate the implications for supersymmetry from an assumed
absence of any signal in the first period of LHC data taking at 7\,TeV
center-of-mass energy and with 1 to 7\,fb$^{-1}$ of integrated
luminosity. We consider the zero-lepton plus four jets and missing
transverse energy signature, and perform a combined fit of low-energy
measurements, the dark matter relic density constraint and potential
LHC exclusions within a minimal supergravity model. A non-observation
of supersymmetry in the first period of LHC data taking would still
allow for an acceptable description of low-energy data and the dark
matter relic density in terms of minimal supergravity models, but
would exclude squarks and gluinos with masses below 1\,TeV.
\end{abstract}


\maketitle

\section{Introduction}
\label{sec:intro}

The Higgs sector of the Standard Model (SM) of elementary particle
physics suffers from the hierarchy problem between the weak scale and
the Planck scale. Extending the SM through supersymmetry (SUSY) is a
very promising solution \cite{Nilles:1983ge}.  The new supersymmetric
particles must then have masses of order 1 TeV. Furthermore,
supersymmetric particles can contribute to low-energy observables via
radiative quantum corrections. In fact, SUSY models provide an
excellent fit to the extensive data, see for example Refs.\,\cite{de
  Austri:2006pe, Allanach:2007qk, Lafaye:2007vs, Buchmueller:2008qe,
  Bechtle:2009ty}. The fits generically prefer a light SUSY spectrum,
typically below 1\,TeV.

The ATLAS and CMS experiments at the Large Hadron Collider (LHC) at
CERN have so far analyzed 35\,pb$^{-1}$ of integrated luminosity at
7~TeV center-of-mass energy. A total integrated luminosity of 1 to
7~fb$^{-1}$ can realistically be achieved at this energy through
2012. These data will significantly extend the search reach for
SUSY~\cite{Atlas-pub,CMS-pub}.

The minimal supersymmetric SM has 124 free parameters. However,
current precision observables and direct search limits only provide
sensitivity to very restricted SUSY models with a small number of
parameters, like minimal supergravity (mSUGRA) \cite{Nilles:1983ge}
which only has 5 free parameters beyond those of the SM. In the
following, we thus focus on mSUGRA models. Specifically, they are
characterized by a common supersymmetric scalar mass $M_0$, a common
gaugino mass $M_{1/2}$, a universal trilinear coupling $A_0$, the
ratio of the two Higgs vacuum expectation values, $\tan\beta$, and the
sign of the Higgs mixing mass parameter,~sign$(\mu)$.

At the LHC, supersymmetry can be searched for in channels with jets,
leptons and missing transverse energy~\footnote{We restrict ourselves
  to conserved R-parity \cite{Nilles:1983ge} or proton hexality
  \cite{Dreiner:2005rd}.  R-parity violation leads to diluted or no
  missing transverse energy \cite{Dreiner:1991pe}.}.  Studies by ATLAS
and CMS reveal that squarks and gluinos with masses up to about
700\,GeV could be discovered at 7\,TeV energy and 1\,fb$^{-1}$
integrated luminosity \cite{Atlas-pub,CMS-pub}.  If no signal is
found, one can significantly constrain supersymmetric models and
exclude squarks and gluinos with masses close to 1~TeV.

To obtain consistent limits on the SUSY parameter space and the
resulting mass spectrum in the absence of a SUSY signal at the LHC,
one needs to combine potential LHC exclusion limits and current low
energy precision observables in a global fit.  In the following we
employ the {\sc Fittino} framework
\cite{Bechtle:2009ty,Bechtle:2004pc} to study such a scenario.

\section{Fit Observables}
\label{sec:obs}

We follow the {\sc Fittino} analysis in Ref.\,\cite{Bechtle:2009ty}
and consider the following set of low-energy observables and existing
collider limits in light of the mSUGRA model: \textit{(i)} rare decays
of B- and K-mesons; \textit{(ii)} the anomalous magnetic moment of the
muon, $a_\mu$; \textit{(iii)} electroweak precision measurements from
LEP, SLC and the Tevatron and the Higgs boson mass limit from LEP; and
\textit{(iv)}~the relic density of cold dark matter in the universe,
$\Omega_\chi$. In contrast to Ref.\,\cite{Bechtle:2009ty}, we employ
the program {\sc HiggsBounds} \cite{Bechtle:2008jh} and not a rigid
Higgs mass limit. We refer to Ref.\,\cite{Bechtle:2009ty} for a
detailed discussion of the low-energy inputs and the collider limits.

At the LHC, the most stringent limits on supersymmetric models with
$R$-parity conservation can be expected from searches in channels with
jets, leptons and missing transverse energy, $E_T^{\rm miss}$. We
follow the analysis presented in Ref.\,\cite{Atlas-pub} and consider
the search channel with four jets, zero leptons and $E_T^{\rm miss}$.
This channel drives the sensitivity, in particular for large
$M_{1/2}$. The selection cuts are
\begin{list}{}{
  \topsep1mm \labelwidth1.6cm \leftmargin0.4cm \labelsep0.2cm \rightmargin0.5cm 
  \parsep0.5ex plus0.2ex minus0.1ex \itemsep0.1ex plus0.0ex }
\item[--] four or more central jets with the 
pseudorapidity $|\eta({\rm jet})| < 2.5$, and with the transverse
momentum $p_T> 100$\,GeV for the leading jet, and $p_T > 40$\,GeV for
the other jets;
\item[--] an opening angle between the transverse momentum of the three leading jets and 
$\vec{p}_T^{\rm\, miss}$ satisfying $\Delta\phi(\vec{p}_T^{\mathrm{\, jet},i}, \vec{p}_T^
{\rm\, miss}) > 0.2$;
\item[--] the missing transverse energy $E_T^{\rm miss}  > 80$\,GeV;
\item[--] the ratio of the missing transverse energy and the 
effective mass satisfying $E_T^{\rm miss}/M_{\rm eff} > 0.2$;
\item[--] the transverse sphericity\,\cite{sphericity} $S_T > 0.2$;
\item[--] no leptons with $p_T > 20$\,GeV.
\end{list}
The effective mass is defined as the scalar sum of the transverse
momenta of all main objects, \textit{i.e.}\
\begin{equation}
M_{\mathrm{eff}}=\sum_{i=1}^{N_{\rm jets}=4} p_T^{\mathrm{jet},i}
+E_T^{\mathrm{miss}}.
\end{equation}

The SM background processes have been described in detail in
Ref.\,\cite{Atlas-pub}. After the cuts listed above, the combined SM
cross section has been estimated to $\sigma_{\rm SM} = 2.42$\,pb at 7
TeV. The signal cross section is dominated by squark and gluino pair
production, $pp \to \tilde q \tilde q^*, \tilde q \tilde q,\tilde q
\tilde g$ and $\tilde g\tilde g$, but all SUSY pair production
processes are included in our numerical analysis. We use {\sc
  Herwig}{\footnotesize ++}~\cite{Bahr:2008pv} in combination with the
parameterised fast detector simulation {\sc
  Delphes}~\cite{Ovyn:2009tx} to obtain the detector response and, in
particular, the shape of the $M_{\rm eff}$ distribution for a given
point in the supersymmetric parameter space~\footnote{A fast
  implementation of the four-jet signal calculation using the methods
  introduced in Ref.~\cite{Dreiner:2010gv} is work in progress.}.  The
signal estimate is normalized to the NLO+NLL QCD prediction for the
inclusive squark and gluino cross sections~\cite{NLLrefs}.

To obtain good sensitivity to a SUSY signal, the full distribution of
$M_{\mathrm{eff}}$ is included in the statistical analysis. We
consider ten bins in the range $0 < M_{\rm eff} < 4$\,TeV and
calculate the $\chi^2$ contribution to the SUSY parameter fit from the
number of signal and background events in each bin of the $M_{\rm
  eff}$ distribution. We define a test statistic $t = -2\ln Q$ with
$Q$ being the likelihood ratio
\begin{equation}
    Q = \prod_{i=1}^{N_{\mathrm{bins}}}
    \frac{\mathcal{L}(\mu_i=s_i+b_i;n_i)}{\mathcal{L}(\mu_i=b_i;n_i)}.
\end{equation}
Here $\mathcal{L}(\mu;n) = \mu^n e^{-\mu} / n!$ is the Poisson
probability to observe $n$ events if $\mu$ are expected. $s_i$ and
$b_i$ are the expected number of signal and background events in bin
$i$, and $n_i$ is the observed event count in this bin.  $s_i$ is a
function of the SUSY parameters, whereas $b_i$ is fixed.  We consider
a signal excluded with 95\% confidence level (CL) if
\begin{equation}
    \mathrm{CL}_{s+b} = \int\limits_{t_{\mathrm{obs}}}^{\infty}
    P_{s+b}(t) \,dt < 0.05\,.
\label{eq:clsb}
\end{equation}
Here $P_{s+b}(t)$ is the probability density function of $t$ assuming
the presence of a signal and $t_{\mathrm{obs}}$ the actually observed
value of $t$.  Uncertainties on the cross-sections are taken into
account by a correlated smearing of the expected event numbers. A
given $\mathrm{CL}_{s+b}$ value can be approximately translated into a
$\chi^2$ contribution using the formula~\cite{Ellis:2007fu}
\begin{equation}
    \chi^2 = 2 [\,\mathrm{erf}^{-1}(1 - 2 \,\mathrm{CL}_{s+b})]^2\,.
\end{equation}
To obtain expected exclusion limits we use the Asimov data set $n_i =
b_i$, $i=1,\dots,N_{\mathrm{bins}}$.

\section{Numerical results}
\label{se:numres}

We present results from a global fit of the mSUGRA model to low-energy
precision observables, the dark matter relic density, existing
collider data from LEP, SLC and the Tevatron, and potential LHC
exclusion limits. We consider LHC scenarios corresponding to
35\,pb$^{-1}$, and 1, 2 and 7\,fb$^{-1}$ integrated luminosity.

\begin{figure}
\includegraphics[scale=0.415]{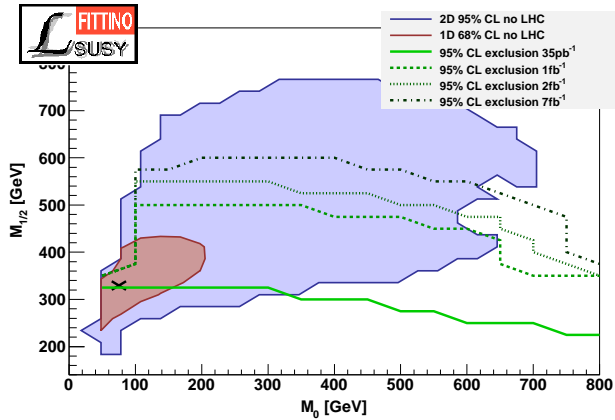}
\caption{mSUGRA parameter region in $M_0$ and $M_{1/2}$ compatible
  with low-energy observables, current collider data from LEP, SLC and
  the Tevatron, and the dark matter relic density. Shown are the
  two-dimensional 95\% and one-dimensional 68\% CL regions. Also shown
  is our estimate of the potential LHC 95\% CL exclusion limits in the
  four-jet, zero-lepton and $E_T^{\rm miss}$ channel for different
  integrated luminosities.}
\label{fig:leo_and_exc}
\end{figure}

In Fig.\,\ref{fig:leo_and_exc} we show the mSUGRA parameter region in
$M_0$ and $M_{1/2}$ compatible with the existing low energy
observables, the existing collider limits from LEP, SLC and the
Tevatron, and the cold dark matter relic density, but no LHC
exclusions imposed.  Note that a positive sign of $\mu$ is preferred
to describe the anomalous magnetic moment of the muon, so we have
fixed sign$(\mu) = +$ and determined $M_0$, $M_{1/2}$, $A_0$, and
$\tan\beta$ from the fit. Including the $1\sigma$ uncertainty, we find
$M_0 = 75\, _{- 29}^{+ 115}$\;GeV, $M_{1/2} = 329 \, _{- 83}^{+
  92}$\;GeV, $A_0 = 417 \, _{- 725}^{+ 715}$\;GeV and $\tan\beta =
13\, _{- 7}^{+ 10}$, in good agreement with
Ref.\,\cite{Bechtle:2009ty}. The minimum $\chi^2$ value is 19 for 20
degrees of freedom. Note that in regions with small $M_0$ and large
$M_{1/2}$ the $\tilde{\tau}$ may be the lightest supersymmetric
particle; such regions are thus excluded from the fit. In
Fig.\,\ref{fig:leo_and_exc} we also show our estimate of the LHC
exclusion limits in the four-jet, zero-lepton and missing transverse
energy channel for different integrated luminosities. The area below
the curves can be excluded at 95\%~CL.  We assign 30\% systematic
uncertainty to the SUSY signal cross-section, and 20\% systematic
uncertainty to the background event rate in the signal-enriched
region, which is dominantly electroweak~\cite{Atlas-pub}.  The LHC
signal estimate for the four jets, zero leptons and $E_T^{\rm miss}$
channel is insensitive to varia\-tions in $\tan\beta$ and $A_0$ within
the systematic uncertainty. Our results are in good agreement with
current LHC limits at 35\,pb$^{-1}$\,\cite{Khachatryan:2011tk,
  atlas_aspen}, and with the projected ATLAS discovery potential at
higher luminosities\,\cite{Atlas-pub}, bearing in mind that we use an
improved signal estimate including the NLO+NLL QCD corrections.

\begin{figure}
\includegraphics[scale=0.415]{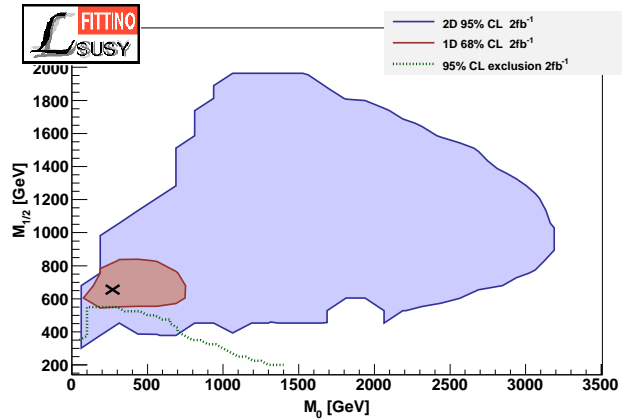}
\caption{mSUGRA parameter region in $M_0$ and $M_{1/2}$ compatible
  with low-energy observables, current collider data from LEP, SLC and
  the Tevatron, the dark matter relic density, and a potential LHC
  exclusion limit in the four-jet, zero-lepton and $E_T^{\rm miss}$
  channel for 2\,fb$^{-1}$ integrated luminosity. Also shown is the
  the potential LHC 95\% CL exclusion limit.}
\vspace*{-2mm}
\label{fig:leo_and_lhc}
\end{figure}

\begin{figure}
\vspace*{4mm}
\includegraphics[scale=0.415]{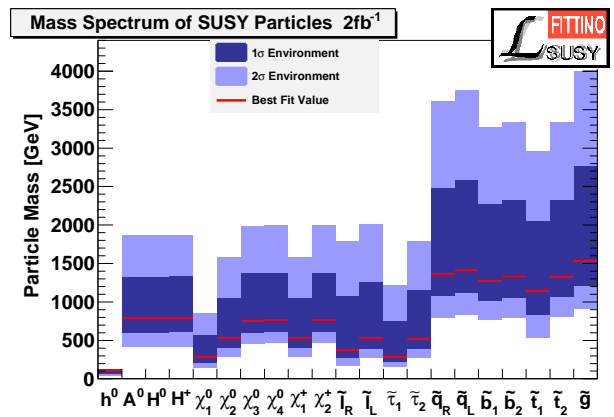}
\caption{SUSY mass spectrum as predicted by a combined mSUGRA fit of
  low-energy observables, current collider data from LEP, SLC and the
  Tevatron, the dark matter relic density, and a potential LHC
  exclusion limit in the four-jet, zero-lepton and $E_T^{\rm miss}$
  channel for 2\,fb$^{-1}$ integrated luminosity.}
\vspace*{-3mm}
\label{fig:leo_and_lhc_mass.eps}
\end{figure}

We now combine the potential LHC exclusion limits, the current
low-energy precision and collider observables, and the dark matter
relic density constraint in a global fit. We assume 2\,fb$^{-1}$
integrated luminosity as our base scenario, but also discuss the
impact of the LHC exclusions at 35\,pb$^{-1}$, and at 1 and
7\,fb$^{-1}$ below. The result of our combined mSUGRA fit assuming no
SUSY signal at the LHC with 2\,fb$^{-1}$ is shown in
Fig.\,\ref{fig:leo_and_lhc}.
The best fit now corresponds to $M_0 = 270\, _{- 143} ^{+ 423}$\;GeV,
$M_{1/2} = 655 \,_{- 81} ^{+ 150}$\;GeV, $A_0 = 763 \, _{- 879} ^{+
  1238}$\;GeV and $\tan\beta = 32 \, _{- 21} ^{+ 18}$, with a minimum
$\chi^2$ value of 24 for 21 degrees of freedom. The corresponding
sparticle mass spectrum is presented in
Fig.\,\ref{fig:leo_and_lhc_mass.eps} and features most probable squark
and gluino masses beyond 1~TeV.

An LHC exclusion in the zero-lepton, four-jet plus
$E_T^{\mathrm{miss}}$ channel is mainly sensitive to the squark and
gluino masses and would drive $M_0$ and $M_{1/2}$ to larger values.
The low-energy precision ob\-ser\-va\-bles and the relic density, on
the other hand, are mainly constraining the masses of colour-neutral
sparticles. Supersym\-me\-tric models with common scalar and gaugino
masses like mSUGRA connect these two, leading to a tension between the
two sets of observables. In addition, for larger $M_0$ and $M_{1/2}$
both $a_{\mu}$ and $\Omega_\chi$ require an increased $\tan\beta$. It
is also noteworthy that the global fit allows areas in the SUSY
parameter space at 95\%~CL, which are located in the region of 95\% CL
exclusion of the LHC, see Fig.\,\ref{fig:leo_and_lhc}. This is due to
the weak dependence of the LHC contribution to the $\chi^2$ on
$M_{1/2}$.  Furthermore values of $M_0$ and $M_{1/2}$ below the direct
LHC limit allow for a significantly better $\chi^2$ from the low
energy data, compensating the contribution from the LHC. Thus the
lower limits on the SUSY masses from the global fit including the LHC
are significantly lower than the direct exclusion limits.

\begin{figure}
\subfigure{\includegraphics[scale=0.20]{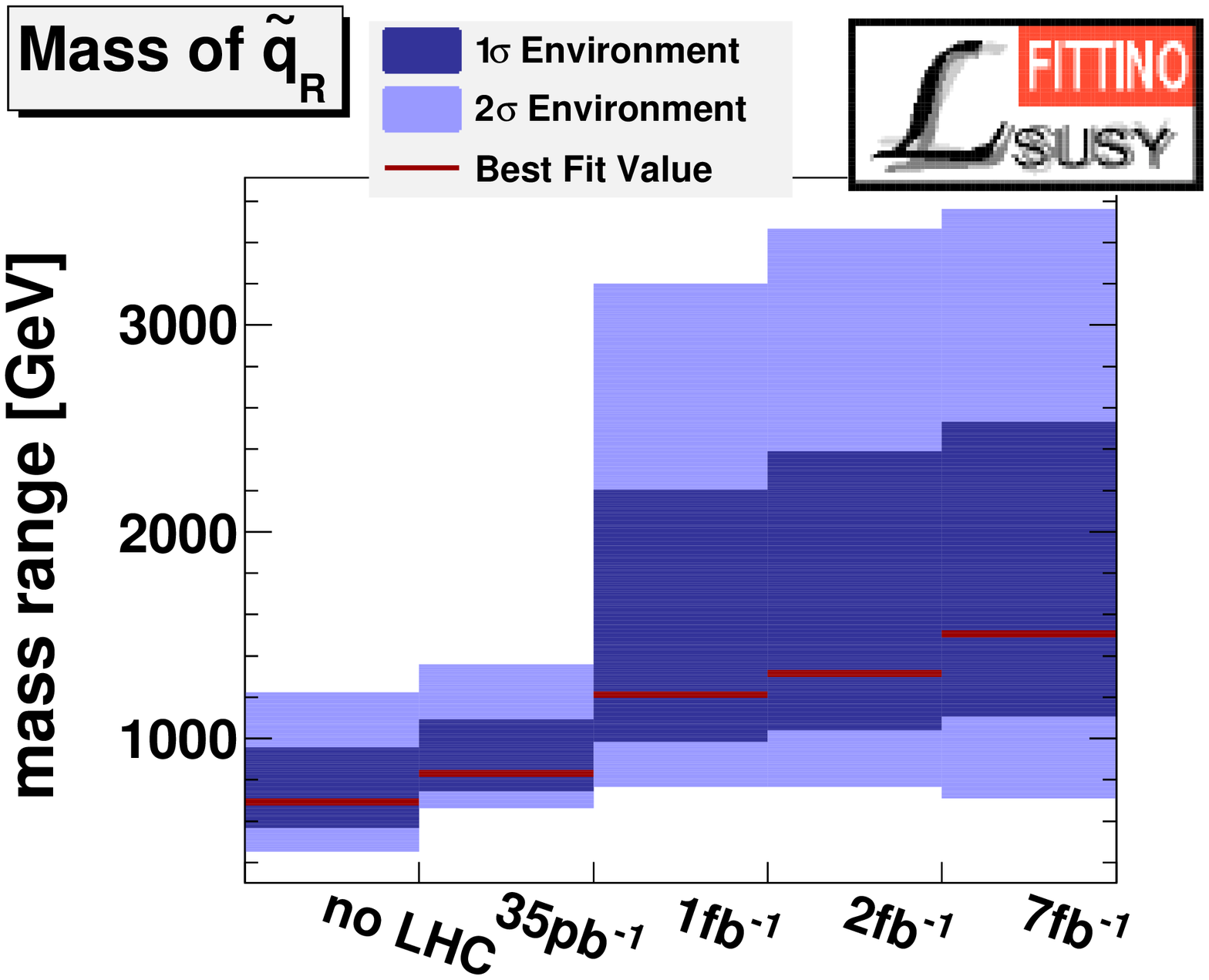}}\hfill
\subfigure{\includegraphics[scale=0.20]{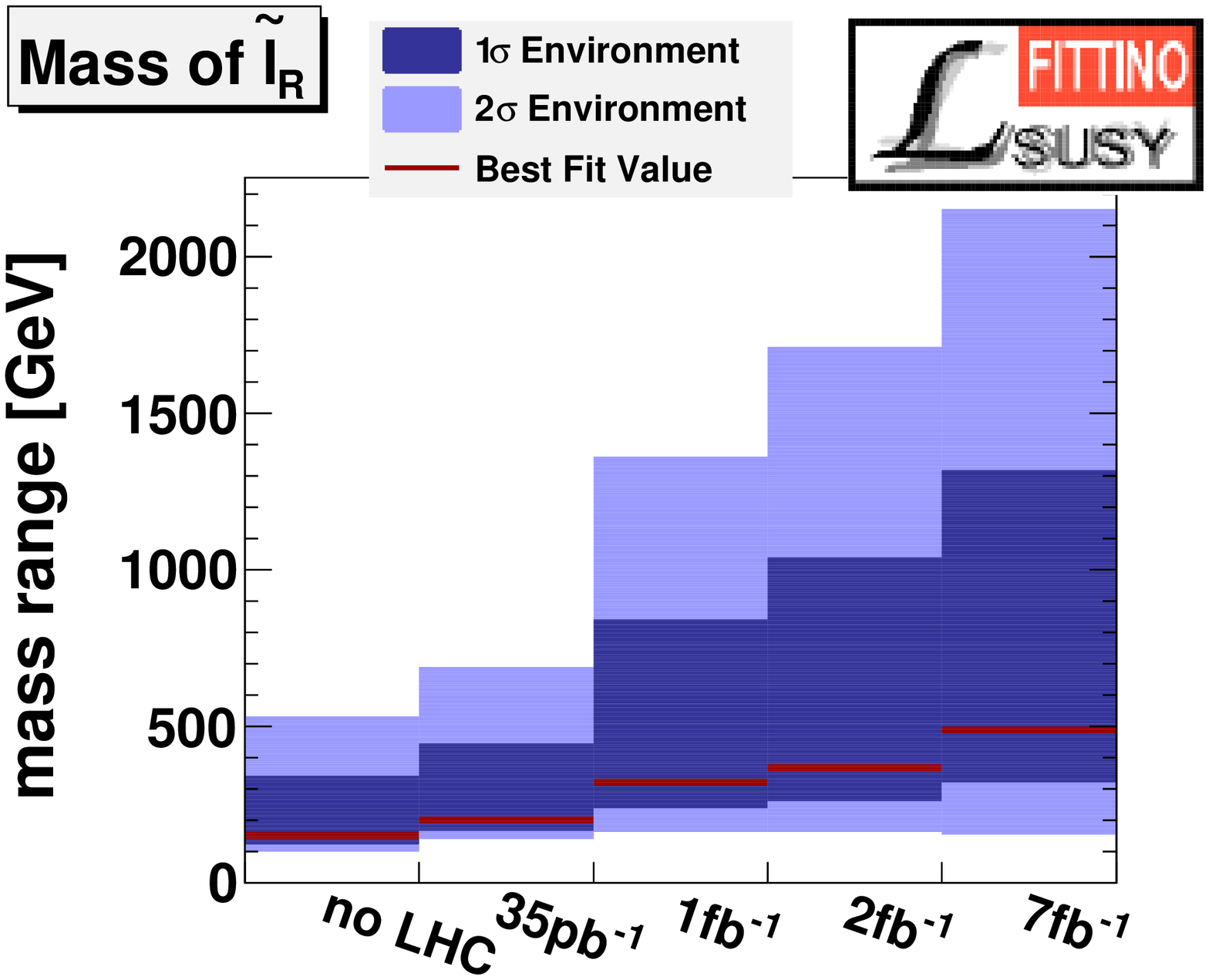}}
\caption{$\tilde{q}_R$ and $\tilde{l}_R$ masses as predicted by a
  combined mSUGRA fit of low-energy observables, current collider data
  from LEP, SLC and the Tevatron, the dark matter relic density, and a
  potential LHC exclusion limit in the four-jet, zero-lepton and
  $E_T^{\rm miss}$ channel for 35\,pb$^{-1}$, and 1, 2 and
  7\,fb$^{-1}$ integrated luminosity. }
\vspace*{-2mm}
\label{fig:leo_and_lhc_squarks}
\end{figure}

Fig.\,\ref{fig:leo_and_lhc_squarks} presents the impact of the LHC
exclusions on the $\tilde{q}_R$ and $\tilde{\l}_R$ mass spectrum from
the global mSUGRA fit, assuming 35\,pb$^{-1}$, and 1, 2 and
7\,fb$^{-1}$.
Already with 1\,fb$^{-1}$ the LHC exclusion would push the lower limit
on the squark mass above the TeV-scale. The $\tilde{\l}_R$ mass, which
is mainly determined from the low-energy observables, is predicted in
the range between 300\,GeV and 500\,GeV. Note that a combined mSUGRA
fit without the constraints on $a_{\mu}$ and $\Omega_\chi$ results in
very large sparticle mass uncertainties.

\section{Conclusions}
\label{se:conclusion}

We have presented a first global mSUGRA analysis of supersymmetric
models which includes the current low-energy precision measurements,
the dark matter relic density as well as potential LHC exclusion
limits from direct SUSY searches in the zero-lepton plus jets and
missing transverse energy channel.

We conclude that it is, in principle, possible to reconcile the
supersymmetric description of low-energy observables and the dark
matter relic density with a non-observation of supersymmetry in the
first phase of the LHC, despite some tension building up in a combined
fit within the mSUGRA framework.  Moreover, we find that a global
mSUGRA fit including potential LHC exclusion limits would yield lower
bounds on squark and gluino masses of about 1\,TeV already with
1\,fb$^{-1}$ integrated LHC luminosity.

While our study is exploratory in the sense that it is based on one
search channel only, and on a simplified description of the LHC
detectors, it clearly demonstrates the potential of the first phase of
LHC running at 7\,TeV in 2011/12 to constrain supersymmetric models
and the sparticle mass spectrum.

\section*{Acknowledgments}
\noindent We thank Sascha Caron and Werner Porod for valuable
discussions. This work has been supported in part by the Helmholtz
Alliance ``Physics at the Terascale'', the DFG SFB/TR9 ``Computational
Particle Physics'', the DFG SFB 676 ``Particles, Strings and the Early
Universe'', the European Community's Marie-Curie Research Training
Network under contract MRTN-CT-2006-035505 ``Tools and Precision
Calculations for Physics Discoveries at Colliders'' and the Helmholtz
Young Investigator Grant VH-NG-303. MK thanks the CERN TH unit for
hospitality.

\bibliographystyle{h-physrev}

\end{document}